# Silver-assisted growth of NdBa$_2$Cu$_3$O$_{7-\delta}$ thin films: A novel approach for the growth of superior quality ceramic oxide films

J. Kurian,[a)] H. Sato, T. Makimoto and M. Naito [b)]

NTT Basic Research Laboratories, NTT Corporation, 3-1 Morinosato-Wakamiya, Atsugi-shi, Kanagawa 243 0198, Japan

Abstract

We have grown NdBa$_2$Cu$_3$O$_{7-\delta}$ films under silver atomic flux by molecular beam epitaxy, which show a drastic improvement in microstructure and also crystallinity leading to 30 % enhancement in critical current density. The most remarkable point is that the final film is *free from* silver. The key to our process in achieving a silver-free film was the use of RF-activated oxygen that oxidizes silver, nonvolatile, to silver oxide, volatile at the deposition temperature. This process enables one to utilize the beneficial effects of silver in the growth of oxide films and at the same time ensures that the final film be free from silver, which is important for high-frequency applications. This method can be made use in the growth of thin films of other complex oxide materials.

[a)] Present address: TU Darmstadt, Petersenstrasse 23, 64287 Darmstadt, Germany

[b)] Present address: Department of Applied Physics, Tokyo University of Agriculture and Technology, Tokyo 184-8588, Japan

There are many reports in the literature about the versatile nature of silver (Ag) in modifying the microstructure of ceramic oxide materials and thereby instituting beneficial improvements in their microstructure and other properties in cuprate superconductors,[1-9] dielectric materials,[10,11] etc. In the case of cuprate superconductors, the films grown by pulsed laser deposition (PLD) from Ag-superconductor composite targets show an improvement in microstructure and an enhancement in critical current density ($J_C$).[2-4,6-9] Some of the reports show an optimum value of Ag in the target is essential to achieve the best results.[2,9] Amid the improvement in microstructure and other physical properties in these films, the final film always contained some amount of Ag.[2,7,9,12-14] Also, it has been found that Ag stays in the metallic form.[7,8,13] This would be undesirable in some cases where the presence of metallic Ag can be detrimental to certain characteristics like high-frequency loss in superconducting films or dielectric loss in dielectric films. Here we report the case of Ag assisted growth of superconducting NdBa$_2$Cu$_3$O$_{7-\delta}$ (NBCO) thin films under RF activated atomic oxygen by molecular beam epitaxy (MBE), where we could reap the benefits of Ag in the improvement in microstructure and $J_C$, and yet the final film is *free from* Ag where Ag has more of a catalytic role. The RF activated atomic oxygen during growth oxidizes the metallic silver to silver oxide, which re-evaporates from the substrate surface at the deposition temperature. This method could be extended to the growth of other oxide ceramic films where silver does not interact with the primary phase.

NBCO films in the present study were grown on (100) MgO substrate by molecular beam epitaxy (MBE) in a custom designed ultra high vacuum chamber with a load-lock chamber arrangement. MgO substrate was used as it is the best available choice for high frequency applications.[15] The MBE chamber is equipped with reflection high energy electron diffraction (RHEED) monitoring set-up for the real time monitoring of the growth process and it enables us to make necessary adjustments during growth, if needed. The films were grown by e-beam coevaporation from Nd, Ba, Cu & Ag metal sources and the precise rate control of the individual species were achieved by feedback controls to the e-guns from sensors utilizing electron impact emission spectrometric (EIES) technique (Sentinel III, Inficon). More details of the growth set-up and the growth of NBCO films are reported elsewhere.[15,16] The oxidation during growth was achieved by RF activated oxygen source (HD60, Oxford) that enables strong and yet controllable oxidation. The pre-annealed MgO (1000 $^o$C for 10 h in oxygen) were heated radiantly and the temperature was controlled by a thermocouple calibrated by an optical pyrometer. The NBCO films were grown at a temperature of ~700 $^o$C under RF activated oxygen with ~2 sccm flow and 300 W RF power. The typical film growth rate was ~25 nm/min. After the growth, the films were cooled below 200 $^o$C in radical oxygen under the same flow rate before removing from the chamber. The films were then subjected to oxygen annealing in an external furnace for at 320 $^o$C for full oxygen loading. The typical thickness of the films in the present study were ~500 nm. The films were characterized by X-ay diffraction, secondary ion mass spectrometry (SIMS), temperature-resistivity measurements, critical current measurements, microwave surface measurements, etc.

One of the distinct characteristics of the NBCO films grown with an Ag atomic flux (we call "Ag-assisted") compared to that of NBCO films grown without Ag atomic flux ("standard") is the marked modification of the microstructure of the films. Figure 1(a) shows the typical atomic force microscope (AFM) (Nanoscope III, Digital Instruments) image of the surface of a standard NBCO film where the usual grain size is between 250 to 400 nm. The AFM image of the surface of an Ag-assisted NBCO films grown with Ag atomic flux rate of ~0.3 mol is shown in Fig. 1(b). The modification of the surface morphology by the effect of Ag is clear from Fig. 1(a) and 1(b). The Ag-assisted NBCO films show an enlargement in grain size with the typical grain size between 1 – 1.5 $\mu$m. Another feature of the surface of Ag-assisted NBCO films is that the surface of the films was much smoother as evident from the root mean square (RMS) surface roughness values. The typical RMS surface roughness values for standard NBCO films grown on MgO (thickness 500 nm) were ~6 to 8 nm (for a 4 $\mu$m × 4 $\mu$m area) whereas for Ag-assisted films was ~2 nm. It is worth mentioning that the AFM images and roughness values of Ag-assisted NBCO films on (100) MgO are close to the ones observed in the case of NBCO films grown on lattice-matched LaAlO$_3$ or SrTiO$_3$ where a 2D layer-by-layer growth was observed.[17-19] Furthermore, one could see shallow growth spirals on the surface in standard NBCO films grown on MgO whereas we could not observe any growth spirals on the surface in Ag-assisted films. Figure 1(c) shows a more enlarged view of the surface of a NBCO film grown under Ag atomic flux and the line profiles along the indicated straight line. It is worth mentioning that the steps observed on the surface of NBCO films grown under an Ag flux were of unit cell dimension and also the faceting of the terraces is clear in the AFM image (Fig. 1(c)). We have studied the growth of NBCO films with different Ag atomic flux rates (corresponding 0 to 1.5 mol) and did not find much difference in results for Ag atomic flux rates above ~0.15 mol. Below this value of Ag atomic flux rate, the modification in microstructure or other properties were not prominent.

The structure of the NBCO films was examined by X-ray diffraction (XRD) using a Rigaku X-ray diffractometer (RINT, Japan) for $2\theta$-$\theta$ scans and a Phillips four circle X-ray diffractometer (X'Pert-MRD) for $\omega$ and $\phi$-scans measurements with Cu-K$_\alpha$ radiation. Figure 2 shows the typical $2\theta$-$\theta$ XRD pattern of 500 nm thick Ag-assisted

NBCO thin film on (100) MgO. In the XRD pattern, all the peaks except the characteristic peak of MgO are that of (00$l$) reflection of NBCO indicating the good $c$-axis texture of the NBCO films. The crystallinity of the NBCO films was examined by employing $\omega$-scan of (006) reflection of NBCO. The full width at half maximum (FWHM) of the rocking curve ($\Delta\omega$) for Ag-assisted NBCO films was ~0.12 degrees whereas $\Delta\omega$ for standard NBCO films was ~0.2 deg. As expected from the AFM results, the assistance of Ag reduces $\Delta\omega$ almost to half. The excellent in-plane orientation of the NBCO films on MgO is evident from the $\phi$-scan measurement of (103) reflection of NBCO (inset of Fig. 2). There was also a reduction in $\Delta\phi$ for Ag-assisted NBCO films grown on MgO (~1.2 deg) compared to that of standard NBCO films (~1.8 deg) indicating an improvement in the in-plane alignment.

In the previous studies where $YBa_2Cu_3O_{7-\delta}$ (YBCO) films were grown by PLD using YBCO-Ag composite targets show the presence of a considerable amount of Ag in the films. The amount of Ag in the film increased with the increase in Ag in the target (the amount of Ag in the films were reported to be lower than in the target) and showed that an optimum level of Ag in the target was essential to achieve the best results.[2,9] Also, some of the studies has established that Ag stays in the metallic form in the final film.[8,13] Even though, the use of Ag improves the microstructure and other properties, the presence of metallic Ag can be detrimental depending on the application. Hence, it is desirable that the final film be free from Ag. To determine whether Ag is present in our NBCO films, we have carried out a SIMS depth profile analysis (CAMECA 1MS-4f, detection area 60 $\mu$m in dia.) of NBCO films grown under different Ag flux rates (nominal Ag flux rates corresponding to 0.3 & 0.9 mol). The SIMS depth profile analysis did not detect Ag in these films. That is, Ag was absent even in the case where the films grown with an Ag atomic flux rate corresponding to 0.9 mol. This conclusion is supported by the RHEED patterns during growth, which were indistinguishable in the growth of NBCO films with and without Ag flux, indicating that Ag is not staying on the surface for 'long' during growth. The absence of Ag in and also on the films implies that the present scenario realized in Ag-assisted NBCO growth differs from those encountered in surfactant-mediated epitaxial growth. It is plausible that as we are using RF activated atomic oxygen for the oxidation during growth, the silver is oxidised to volatile silver oxide which re-evaporates from the surface of the substrate at the deposition temperature. The highly volatile nature of silver oxide was verified by the following experiment. We have evaporated Ag alone on MgO substrate with RF activated oxygen at the same deposition temperatures for NBCO growth. Ag was not observed on the surface of MgO (observed under microscope). On the other hand, with molecular oxygen (simply RF switched off), metallic Ag was observed on the surface of MgO.

The superconductivity of the NBCO films was studied by standard four-probe technique. Figure 3 shows the typical variation of resistivity with temperature for an Ag-assisted NBCO film, which shows good metallic behavior in the normal state ($\rho_{300K}/\rho_{100K}$ ~3) and gave a $T_C(0)$ above 94 K with a sharp transition ($\Delta$T ~1 K). In fact, the $\rho$-$T$ curves and values were not much different from those of standard NBCO films. However, the critical current density ($J_C$) determined by induction method for the Ag-assisted NBCO films shows some enhancement (20 to 30%), typically giving ~ 4.5 MA/cm$^2$ at 77 K, compared with $J_C$ ~ 3.5 MA/cm$^2$ for standard films. The $J_C$ values of ~ 4.5 MA/cm$^2$ are fair high values for NBCO films on bare (100) MgO. The $J_C$ values of Ag-assisted NBCO films grown with different Ag atomic flux rates (corresponding to 0.3 to 1.8 mol) were all the same. We should note that the rather moderate enhancement $J_c$ in spite of the significant improvement in microstructure partly comes from the difficulty in oxygen loading due to large grain size as encountered in liquid-phase epitaxy (LPE) films.

The microwave surface resistance ($R_S$) of NBCO films was measured by dielectric resonator method with sapphire rod and an impedance analyzer (Agilent 8510C). A pair of 20 mm $\times$ 20 mm size NBCO films was used for the $R_S$ measurements. Figure 4 shows the temperature dependence of microwave $R_S$ of an Ag-assisted NBCO film measured at 22 GHz. The $R_S$ for Ag-assisted NBCO films was essentially the same as the one for standard NBCO. However, the reproducibility of the $R_S$ values for different runs was much improved by Ag-assisted growth. The low microwave $R_S$ confirms again that normal metal Ag is not left in Ag-assisted NBCO films. The low microwave $R_S$ with an improvement in $J_C$ of Ag-assisted NBCO films should be beneficial in realizing microwave devices with high power handling capability.

In summary, we report the growth of NBCO films with improved characteristics with Ag assisted growth by MBE. The most significant highlight of the present study is that we could utilize the beneficial aspects of Ag in the improvement in microstructure, crystallinity, and $J_{C,}$ and yet the final film is *free from* Ag. Mechanisms for our observations are not yet clear and worth to future microscopic and theoretical studies. At present, we speculate that the highly unstable and volatile nature of silver oxide may be a key to understand our observations. The peculiar nature of silver oxide provides a unique stage for thin film growth with highly enhanced migration of adatoms, involving rapid surface diffusion of Ag, dissociation and recombination of Ag-O bonds, re-evaporation of Ag oxide, and possibly incorporation of Ag in the 123 lattice. Such a stage has been rarely encountered in conventional crystal growth of thin films. Finally we also suggest that Ag-assisted deposition could be made use in the growth of other complex oxide films.

**Acknowledgements**

The authors like to acknowledge H. Shibata, H. Yamamoto, S. Karimoto, K. Ueda and A. Tsukada of NTT Basic Research Laboratories for their valuable suggestions and discussions at various stages of this work. The authors like to thank Prof. L. Alff, TU Darmstadt, Germany for fruitful discussions.

**Figure captions**

Fig. 1. The AFM image of surface of 500 nm thick NBCO films grown on MgO substrate (a) without Ag flux [4 $\mu$m × 4 $\mu$m], and under Ag atomic flux (b) [4 $\mu$m × 4 $\mu$m], & (c) [1 $\mu$m × 1 $\mu$m] with line profile along the indicated straight line.

Fig. 2. The $2\theta$-$\theta$ XRD pattern of NBCO film grown on (100) MgO under Ag atomic flux (in-set shows the $\phi$-scan XRD pattern (103) reflection of NBCO film).

Fig. 3. Typical temperature-resistivity curve of NBCO film grown on MgO under Ag flux and the in-set shows the enlarged portion of the superconducting transition.

Fig. 4. Variation of microwave $R_S$ with temperature of a 500 nm thick NBCO film grown on MgO under Ag atomic flux by MBE at 22 GHz.

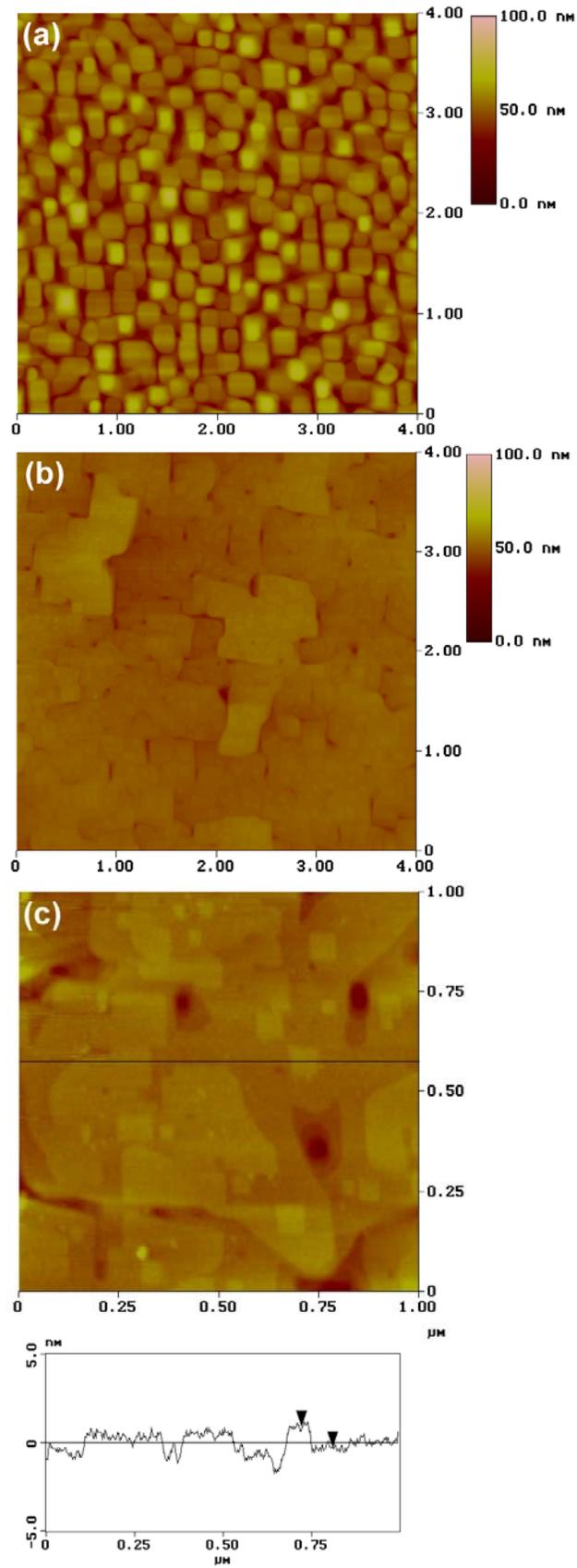

Fig. 1

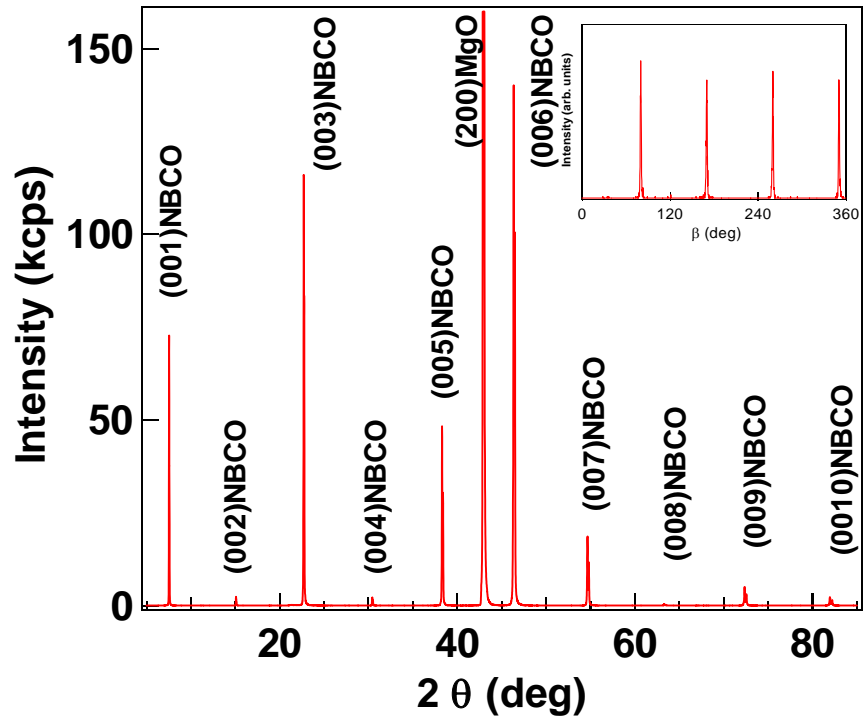

Fig. 2.

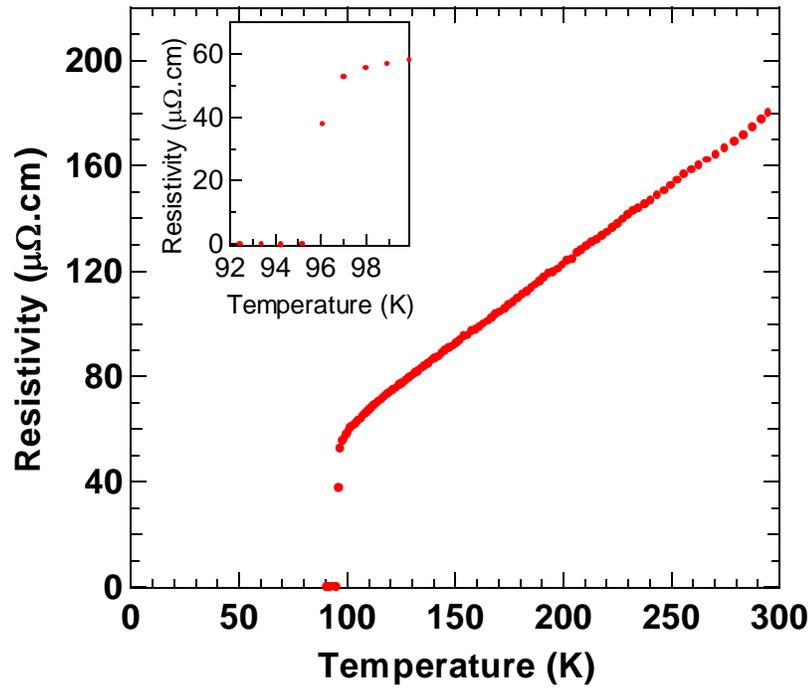

Fig. 3.

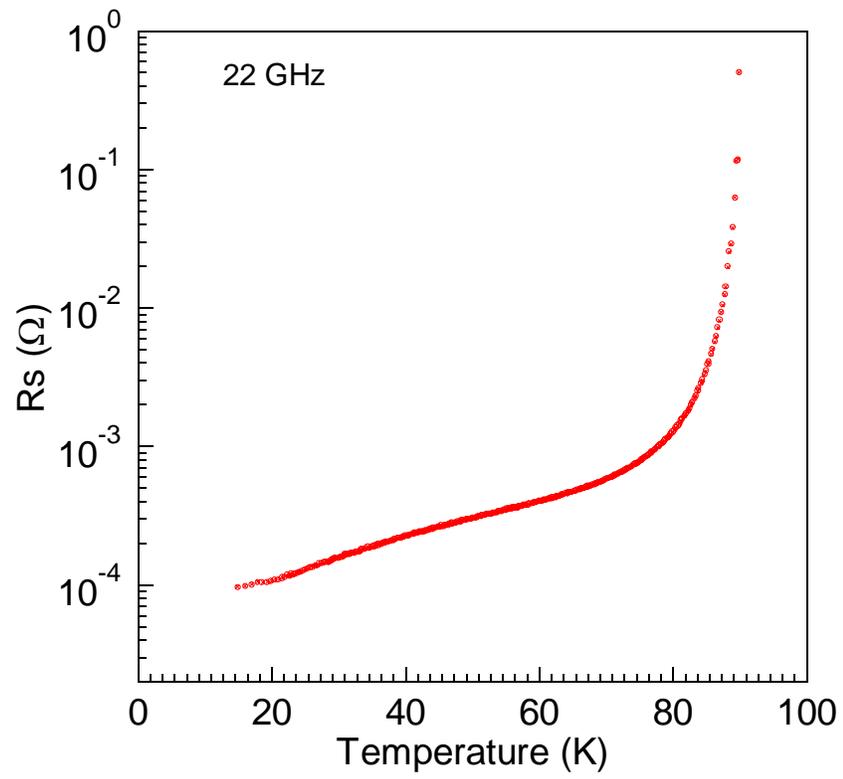

Fig. 4.